\newif\ifcheckpagelimits
\checkpagelimitstrue
\checkpagelimitsfalse

\ifcheckpagelimits
 \documentclass[nofootinbib,prl,aps,twocolumn,showpacs,showkeys,%
 amsmath,amssymb,superscriptaddress,reprint,floatfix,longbibliography]{revtex4-1}
\else
 \documentclass[prl,aps,twocolumn,showpacs,showkeys,%
 amsmath,amssymb,superscriptaddress,reprint,floatfix,longbibliography]{revtex4-1}
\fi

\usepackage{graphicx}
\usepackage{color}
\newcommand{\be}{\begin{equation}}
\newcommand{\ee}{\end{equation}}
\newcommand{\bea}{\begin{eqnarray}}
\newcommand{\eea}{\end{eqnarray}}

\usepackage{hyperref}
\hypersetup{
bookmarks=false,
pdftitle={Chaotic dimer},
anchorcolor=blue,
colorlinks=true,
citecolor=blue,
urlcolor=blue,
linkcolor=blue}
\usepackage{marginnote}
\usepackage[author=??]{pdfcomment}
\usepackage{xcolor}
\ifcheckpagelimits
\newcommand{\mycomment}[1]{}
\else
\newcommand{\mycomment}[1]{\marginpar{\pdfcomment[color=pink,icon=Note]{#1}}}
\fi

\begin{document}

\title{Quantum chaos in an ultra-strongly coupled bosonic junction}

\author{Uta Naether}
\email{naether@unizar.es}
\affiliation{Instituto de Ciencia de Materiales de Arag\'on y Departamento de F\'{i}sica de la Materia Condensada, CSIC-Universidad de Zaragoza, 50009 Zaragoza, Spain}
\author{Juan Jos\'e Garc\'{i}a-Ripoll}
\affiliation{Instituto de F\'{i}sica Fundamental, IFF-CSIC, Serrano 113-bis, 28006 Madrid, Spain}
\author{Juan Jos\'e Mazo}
\affiliation{Instituto de Ciencia de Materiales de Arag\'on y Departamento de F\'{i}sica de la Materia Condensada, CSIC-Universidad de Zaragoza, 50009 Zaragoza, Spain}
\author{David Zueco}
\affiliation{Instituto de Ciencia de Materiales de Arag\'on y Departamento de F\'{i}sica de la Materia Condensada, CSIC-Universidad de Zaragoza, 50009 Zaragoza, Spain}
\affiliation{Fundaci\'{o}n ARAID, Paseo Mar\'{i}a Agust\'{i}n 36, 50004 Zaragoza, Spain}

\date{\today}

\begin{abstract}
The semi-classical and quantum dynamics of two ultra-strongly coupled nonlinear resonators cannot be explained using the Discrete Nonlinear Schr\"odinger Equation or the Bose-Hubbard model, respectively. Instead, a model beyond the Rotating Wave Approximation must be studied. 
In the semi-classical limit this model is not integrable and becomes chaotic for a finite window of parameters. 
For the quantum dimer we find corresponding regions of stability and chaos. 
The more striking consequence for both semi-classical and quantum chaos is that the {\it tunneling} time between the sites becomes unpredictable. 
These results, including the transition to chaos, can be tested in experiments with superconducting microwave resonators.
\end{abstract}

  

\ifcheckpagelimits
\else
\maketitle
\fi


The  unquestionable relevance of  the Discrete Nonlinear Schr\"odinger equation (DNLS) extends beyond the theoretical characterization of waves in nonlinear media \ \cite{scott,reps}, to describe also a rich variety of physical phenomena ranging from biological physics\ \cite{davydov} to gravitational analogues\ \cite{gravdnls}.
It is therefore not surprising that the theoretical predictions of the existence of,  e.g. discrete solitons\ \cite{theosol}, vortices\ \cite{vort1} and self-trapping\ \cite{sant} can be experimentally tested in a wide variety of setups, such as arrays of optical waveguides\ \cite{solex1}, polaritons\ \cite{stpol} or Bose-Einstein condensates\ \cite{obert1}.
In this respect, the DNLS equation bridges nonlinear science and quantum many body physics, since it is the {\it semi-classical} or {\it many boson} limit of the Bose-Hubbard (BH) model, describing amongst other things tunneling of ultra-cold atoms in optical lattices.

The most basic realization of the DNLS is the dimer, formed by two {\it weakly} coupled nonlinear resonators. The two-site DNLS is integrable and exhibits a transition\ \cite{kenkre} from linear oscillatory dynamics (Rabi regime) to self-trapping (localization). Between these limiting cases lays the Josephson regime, where the quantum equivalent of the dimer, the two-site BH model\ \cite{2sitebh} behaves as a bosonic Josephson junction\ \cite{obert2}. Furthermore, the signatures of the classical symmetry breaking bifurcation can also be observed in the quantum limit\ \cite{bif}.
%
%
In that sense the dynamical behavior in the DNLS  resembles the Mott-Superfluid transition in the BH model\ \cite{greiner}.

In this work we study a more general model of two oscillators with amplitude $\psi$ with coupling no longer captured by perturbation theory: the {\it ultra-strongly coupled} (USC) bosonic junction
\ifcheckpagelimits\else
\begin{equation}
\label{dnls}
\dot \psi_k = -i \omega \psi_k + i J(\psi_{1-k} + \theta \,\psi_{1-k}^* ) - i \tilde{\gamma} |\psi_k|^2 \psi_k, \end{equation}
\fi
for $k=0,1$. %
The USC model ( $\theta=1$ above) becomes the DNLS [$\theta=0$ in \eqref{dnls}] with nonlinearity strength $\tilde{\gamma}$ in the limit of weak couplings $|J/\omega|\ll 1$, by means of the Rotating Wave Approximation (RWA). However, in the ultra-strong coupling regime, where $|J|\simeq |\omega|/2$ (see suppl. mat.), the RWA breaks down and new physics is found. The quantum equivalent of\ \eqref{dnls} are now the ultra-strongly coupled nonlinear resonators,
\ifcheckpagelimits\else
\begin{equation}
\label{H}
H = 
\sum_{k=0,1}
\left [ \omega \hat{a}_k^\dagger \hat{a}_k 
+
\frac{\tilde\gamma}{2} (\hat{a}_k^\dagger)^2 \hat{a}_k^2\right]
-
J (\hat{a}_0^\dagger \hat{a}_1 + \theta\, \hat{a}_0^\dagger \hat{a}_1^\dagger + \mathrm{H.c.}),
\end{equation}
\fi
with Fock operators $[\hat{a}_k,\hat{a}_j^\dagger]=\delta_{jk}$ of both oscillators with frequency $\omega$. This quantum dimer lacks a superselection rule and no longer conserves the number of particles, $\hat{N}=\sum_k \hat{a}^\dagger_k \hat{a}_k$, just like the classical model\ \eqref{dnls} no longer conserves the norm $N=\sum_k|\psi_k|^2$, when $\theta=1$.

The semi-classical and quantum versions of these ultra-strongly coupled bosonic Junctions are of great relevance in the study of superconducting quantum circuits\ \cite{You2011}. Labeled as {\it quantum optics on a chip}, quantum circuits have reproduced most interesting features of cavity quantum electrodynamics (QED) using photons and superconducting qubits as simulators of light and atoms. One of their greatest advantage is the possibility of pushing the light-matter interaction strength close to the energy of the bare frequency transitions ---the ultra-strong coupling.  Quite recently this new regime has been
experimentally demonstrated in these and other solid state setups\ \cite{us}.
Along this letter the ultrastrong refers to the coupling between 
two nonlinear bosonic modes.  
Couplings between superconducting resonators, as the ones we are going to discuss in this letter, recently has been reported experimentally \cite{Haeberlein2013}.

What Physics can be expected in these new coupling regimes? Our study of the ultra-strongly coupled semi-classical and quantum bosonic dimers reveals that both systems experience a transition to chaos for negative values of $\tilde\gamma$. The chaotic regions are finite and can be characterized both spectrally and in phase space, but the most clear signature is the change in the self-trapping dynamics, found in the unpredictability of the tunneling time. These features can be observed using quantum circuits, either in the few-photon or in the semiclassical regime.

\paragraph{The semi-classical limit.---}
In the limit of large number of excitations, $\hat{n}_k := \langle \hat{a}_k^\dagger \hat{a}_k \rangle \gg 1$, where quantum fluctuations become small, the semiclassical dynamics of the coupled resonators can be approximated using coherent states. Replacing the mean-field values $\hat{a}_k \to \langle \hat{a}_k \rangle :=\psi_k$ in Eq.\ \eqref{H}, we obtain a classical Hamiltonian for the dynamical variables $\psi_k$ which, by the Hamilton equations $\dot \psi_k =-i \partial_{\psi_k^*} \bar H$, evolve according to the ultra-strongly coupled bosonic junction equation\ \eqref{dnls}.

In the DNLS limit ($\theta = 0$) the system is integrable and both the energy and the total number of excitations
\ifcheckpagelimits\else
\begin{equation}
N = \langle \hat{N}\rangle = |\psi_0|^2+|\psi_1|^2 =: \hat{n}_0 + \hat{n}_1.
\end{equation}
\fi
are conserved. Moreover, the equations are symmetric under the transformation $\tilde\gamma\rightarrow-\tilde\gamma$, $\psi\rightarrow\psi^*$ and $\psi_k\rightarrow\psi_k\exp(ik\pi)$ (or $J\rightarrow-J$). This symmetry and conservation law disappear when we consider the USC model, $\theta=1$, which is no longer integrable. This has dramatic consequences for the dynamics.

\begin{figure}[t]
\centering
\includegraphics[width=0.45\textwidth]{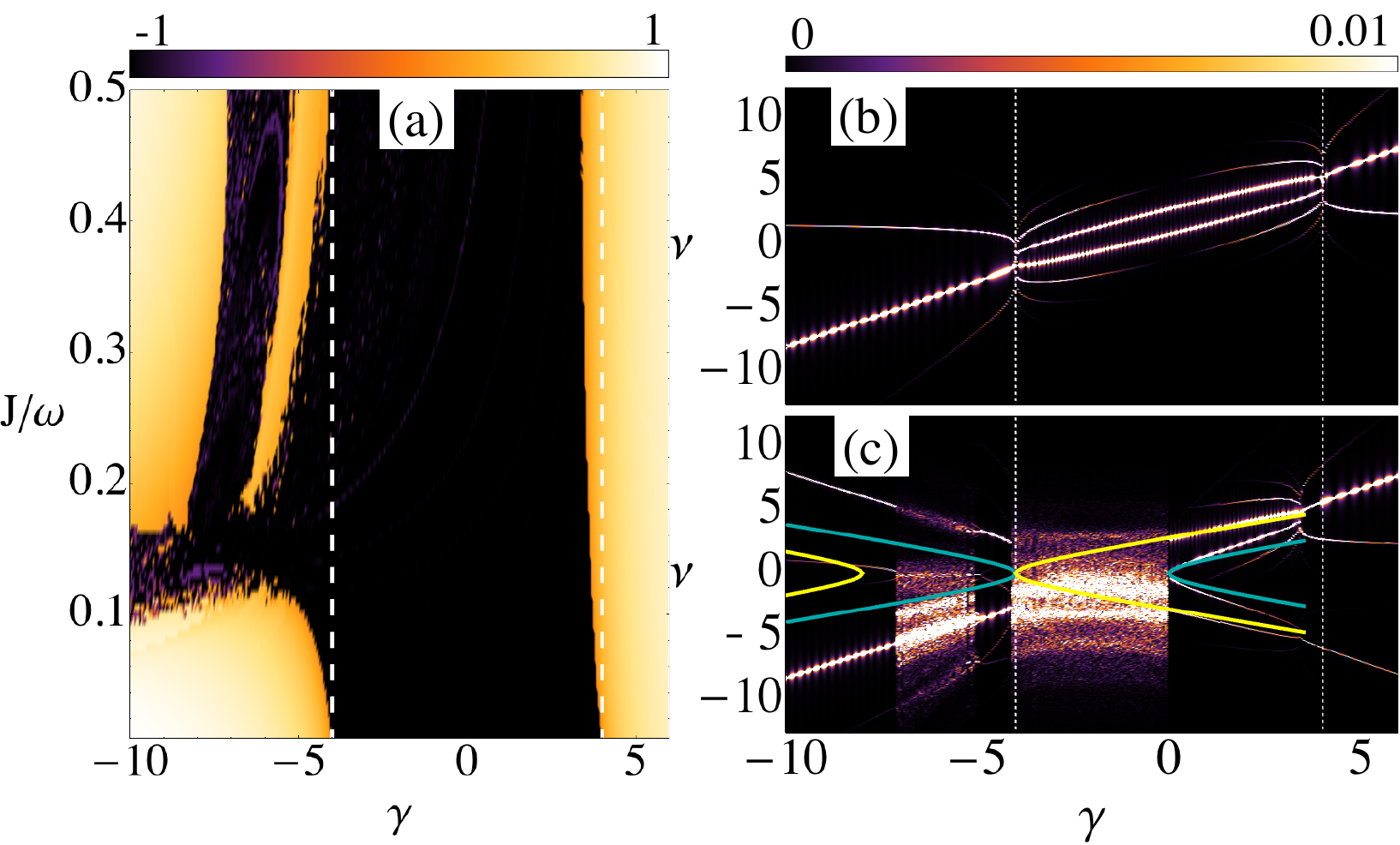}
\caption{(Color Online)(a) $\rho_{min}$ vs. $\gamma$ and $J/\omega$. Dashed white lines stands for the DNLS self-trapping transition, $|\gamma_{c}^{\textsc{\tiny RWA}}|=4$. Spectral densities $g(\nu,\gamma)$ at $J/\omega=0.5$ for the RWA (b), and the USC model (c). The thick lines in (c) shows the analytic continuations of Eq.~\eqref{cont} for $J/\omega=0.5$ and $N_0=1$. Gray (blue) lines are for the symmetric modes and light gray (yellow) lines for the antisymmetric ones.}
\label{fig1}
\end{figure}


We have calculated the semiclassical evolution of the population imbalance $\rho(t) := n_0(t) -n_1(t)$ using Eq.~\eqref{dnls}, comparing the results obtained from the DNLS with the USC model. Along this work we set $J=1$ and start with the initial condition $n_0(0)=N_0,\, n_1(0)=0$. This leaves as only free parameters the relative coupling $J/\omega$ and interaction strengths $\gamma = \tilde{\gamma} N_0/J$. The normalized minimal imbalance $\rho_{min}=\min_t[\rho(t)/N(t)]$ is a witness of self-trapping: a value $\rho_{min}\simeq -1$ indicates an oscillating dynamics where particles eventually tunneled to the opposite site, while the imbalance remains locked around $\rho_{min}\simeq 1$, when self-trapping dominates the dynamics.
In Fig.\ \ref{fig1}(a) we plot $\rho_{min}$ for the USC model, after a sufficiently long time $t \sim 100/J$, starting with $N_0=1$.  For weak coupling $J/\omega\rightarrow 0$, the DNLS dimer is recovered and the self-trapping transition happens at the analytic value $|\gamma_{c}^{\textsc{\tiny RWA}}|=4$\ \cite{kenkre}, denoted by white dashed lines in Figs.~\ref{fig1}(a-c). For increasing coupling strength, $J/\omega$, and positive $\gamma$, self-trapping is observed at slightly smaller nonlinearities $\gamma$. More interesting is the behavior for negative $\gamma$. Increasing $J/\omega$ from zero, the transition is shifted to values $|\gamma|>|\gamma_{c}^{\textsc{\tiny RWA}}|$, reaching a minimum at $J/\omega\approx0.1$. Around this value we begin to observe strong irregular oscillations of $\rho_{min}$.

To better understand the dynamics in this parameter region we compute the normalized spectral density
\ifcheckpagelimits\else
\begin{equation}
\label{g}
g (\nu;\gamma) := \frac{|f_0(\nu;\gamma)|^2 + |f_1(\nu;\gamma)|^2 } {\sum_{\nu} \left[ \, |f_0(\nu;\gamma)|^2 + |f_1(\nu;\gamma)|^2 \, \right]},
\end{equation}
\fi
defined in terms of $f_k(\nu)$, the Fourier transform of $\psi_k(t)$.
Figs.\ \ref{fig1}(b) and (c) show $g(\nu;\gamma)$ for the RWA case and the full mode respectively, at fixed $J/\omega=0.5$, with dashed vertical lines delimiting the analytical prediction $|\gamma_{c}^{\textsc{\tiny RWA}}|=4$. In Fig.\ \ref{fig1}(b) we see uniform and well separated lines, indicating that the dynamics is dominated by only a few frequencies, which we identify with the nonlinearly shifted normal modes or the decoupled localized mode.  However, in Fig.\ \ref{fig1}(c) we observe two broad windows with a huge number of frequencies involved in the dynamics, indicating the presence of chaos for negative values of $\gamma$. 

\begin{figure}[t]
\centering
\includegraphics[width=0.45\textwidth]{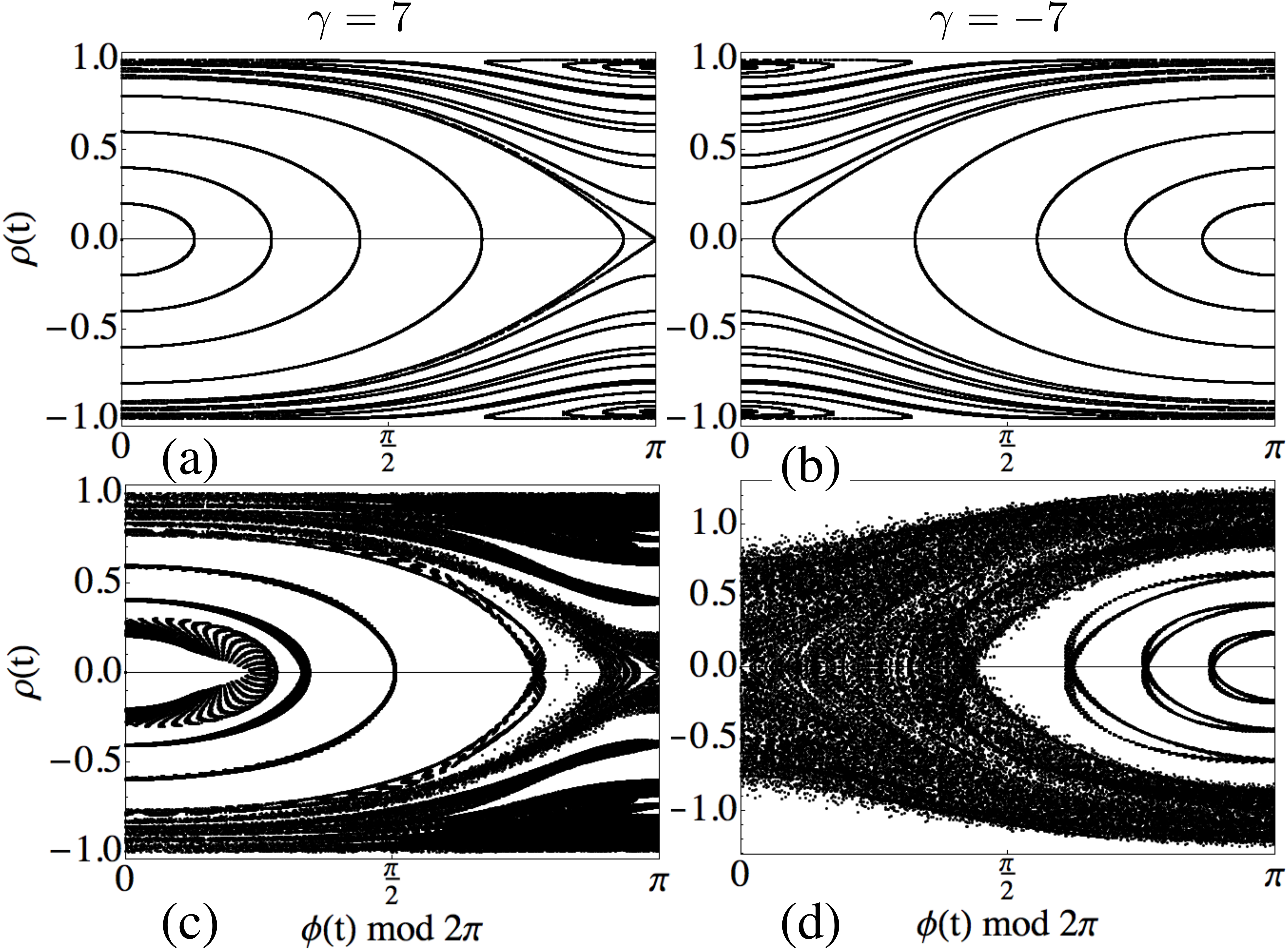}
\caption{Phase space diagrams (top with $\theta=0$) and Poincar\'e sections (bottom for $\theta=1$) with $J/\omega=0.5$, (a),(c): $\gamma=7$ with $\rho(0)\in(-1,1)$, $\phi(0)=0$  and (b),(d): $\gamma=-7$, $\rho(0)\in(-1,1)$, $\phi(0)=\pi$ respectively.}
\label{fig2}
\end{figure}

To get an insight about why chaos can be only found for negative $\gamma$, it is worth looking at the linear modes of the systems and their nonlinear continuations. First, we search for linear solutions ($\gamma=0$) of Eq.~\eqref{dnls}, and we find the corresponding eigenvalues $\left\{\nu\right\}$ and eigenvectors of the system (see suppl. mat.). Two modes are symmetric and the other two are anti-symmetric, each having a positive and a negative eigenfrequency. For finite $\gamma$, we are able to continue both types of states into the nonlinear regime, finding that
\ifcheckpagelimits\else
\bea
\nu_{\uparrow\uparrow}&=&\pm\sqrt{[\omega+\gamma J/2-J(1-\theta)][\omega+\gamma J/2-J(1+\theta)]}\nonumber\\
&&\label{cont}\\
\nu_{\uparrow\downarrow}&=&\pm\sqrt{[\omega+\gamma J/2+J(1-\theta)][\omega+\gamma J /2+J(1+\theta)]}\nonumber.
\eea
\fi
Since $\nu$ has to be real, in the USC model  ($\theta = 1$) there exists a window in parameter space where one of these modes exists: 
\ifcheckpagelimits\else
\be-\left(\frac{2\omega}{J}+4\right)<\gamma<\left(4-\frac{2\omega}{J}\right)\label{concon}
\ee
\fi
In Fig.\ \ref{fig1}(c) we also plot the eigenfrequencies given by Eq.~\eqref{cont} for $J/\omega=0.5$, $\theta=1$ and $N_0=1$. For positive $\gamma$ all four states exists and trajectories, localized or not, are regular. At negative values of the nonlinear parameter the situation is more complex. For $-4 \! < \! \gamma \!< \!0$, the periodic orbits corresponding to the symmetric states disappear and chaotic trajectories are found. This region corresponds to the first window of chaos, and the chaotic trajectories are found to be dominant till the bifurcation point $\gamma=-4$, where the symmetric states reappear.  
However, for $-8 \! < \! \gamma \! < \! -4$ the antisymmetric modes are missing, and a second region of chaos appears. In Fig.\ \ref{fig1}(a) and (c) it is possible to observe for $\gamma<-4$, first an excitation of the nonlinear self trapped state, and then a weaker chaotic region.   

To obtain a more complete characterization of chaos we have computed Poincar\'e sections 
in the $\rho$-$\phi$ reduced phase space of the system, where $\phi(t):={\rm arg}[\psi_0]-{\rm arg}[\psi_1]$ is the phase difference between oscillators. We plot points for which $N=\langle N\rangle$ coincides with the norm averaged over the integration interval. We start in Figs.\ \ref{fig2}(a-b) with the DNLS or RWA ($\theta=0$), for which the norm $N(t)=\langle N\rangle $ is conserved, and we have two types stationary orbits depending on the value of the initial imbalance: periodic orbits and self trapped states. The situation is quite different in the USC model, where $N$ is not conserved. For positive values of $\gamma$, the Poincar\'e sections show deformed tori corresponding to the quasiperiodic motion of the oscillators~[Fig.\ \ref{fig2}(c)], but for negative $\gamma$, these tori coexist with chaotic trajectories\ [Fig.\ \ref{fig2}(d)].  The chaotic nature of these orbits was confirmed by the computation of Lyapunov exponents, which were found to be positive for the chaotic trajectories shown in Fig.\ \ref{fig2}(d).
\begin{figure}[t]
\centering
\includegraphics[width=0.5\textwidth]{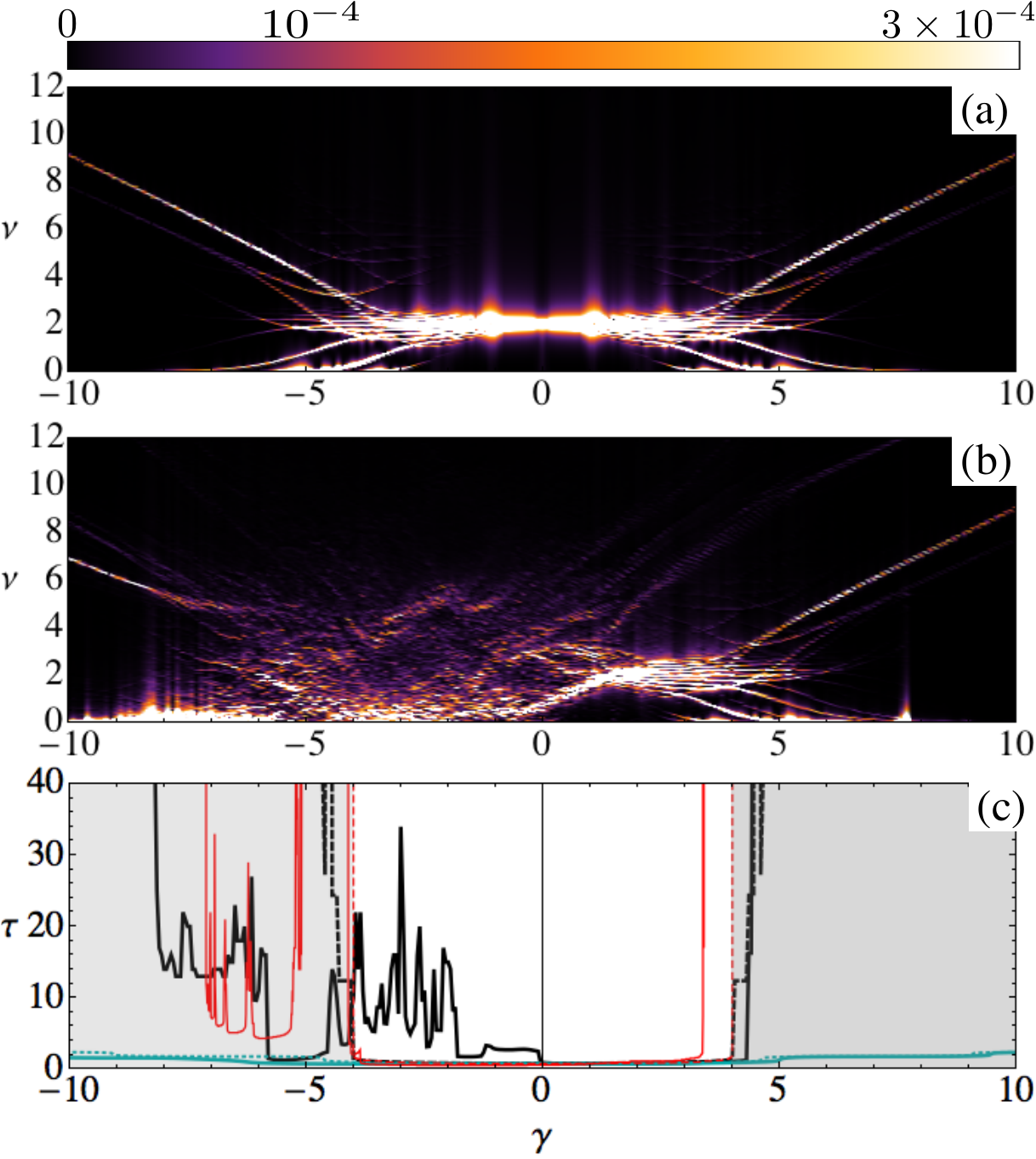}
\caption{(Color online) Quantum dynamics. Spectral density $g(\nu,\gamma)$ for $\theta=0$ (a) and  $\theta=1$ (b) [$N_0=\rho_0=17$ and $\omega=2$]. Figure (c) shows $\tau$ vs. $\gamma N$, for $N_0=\rho_0=17$ (black) and $N_0=\rho_0=2$ shown in green. Full (dashed) lines correspond to  $\theta=1$ ($\theta=0$), respectively. Red lines correspond to the semiclassical model results.}
\label{fig3}
\end{figure}



\paragraph{Quantum dynamics.---}
A similar study has been done for the quantum models in Eq.\ \eqref{H}. Starting with an initial state that corresponds to an imbalanced Fock state $|\psi\rangle = |n_0(0)\rangle \otimes |n_1(0)\rangle$, with $N_0=n_0(0)+n_1(0)$, we simulated the evolution of the number of photons, $n_{0,1}(t)$, and their imbalance, using exact diagonalizations in a truncated Fock basis.

Our first tool for analyzing the dynamics is again the spectral density $g(\nu; \gamma)$ from Eq.\ \eqref{g}, which  is computed replacing $f_k(\nu;\gamma)$ with the Fourier transform of $n_k(t)$ for a given $\gamma$. The spectral density is shown in Fig.\ \ref{fig3} for positive frequencies\ \cite{footnote}. In the quantum case with the RWA ($\theta=0$) we observe the same symmetry $\gamma \to -\gamma$ [cf. Fig.\ \ref{fig3}(a)] as in the classical DNLS dynamics.  The absence of this symmetry in the USC model ($\theta=1$) also manifests in Fig.\ \ref{fig3}(b) with the appearance of a multitude of frequencies for negative $\gamma$, in stark resemblance of our semiclassical signature of quantum chaos.

As in the semiclassical limit [cf. Fig.\ \ref{fig1}] the self-trapping transition for $\gamma<0$ is shifted due to the excitation of chaotic trajectories.  To resolve this transition more clearly, we have computed a dimensionless tunneling time $\tau:=\min(Jt:\rho=0)$ defined as the time at which the population imbalance first changes its sign, for initial conditions $n_0(0)=N_0$, $n_1(0)=0$. Let us first discuss the situation with a large number of particles, $N_0= 17$, where the quantum and semiclassical models are expected to converge. The classical self-trapping regions correspond to the shaded areas in Fig.\ \ref{fig3}(c), and the corresponding semiclassical (quantum) {\em tunneling} times are shown in red (black) lines, either dashed ($\theta=0$, RWA) or solid ($\theta=1$, USC model). For positive $\gamma$ we observe that both in the RWA case and the USC model $\tau$ grows steeply when entering the shaded region. The only difference is that the quantum model does not lower the self-trapping transition. For negative $\gamma$ the RWA curve behaves symmetrically, whereas in the USC model we find an {\it irregular} behavior. Furthermore, we also observe a small window of regularity in the quantum curve, which corresponds to the same phenomenon of a small intermediate region of quasi periodicity observed in the semiclassical case. 

For smaller number of particles, such as $N_0=2$ (green curves), quantum fluctuations become relevant and we enter the so-called {\it Rabi regime} of the dimer\ \cite{obert2}. The first consequence is that self trapping is possible only for very large onsite interaction $\gamma$, outside the range of values considered in this work. The second consequence, the smoothness of the curves, is a clear signature of the lack of chaos. For $N_0=2$   the system is certainly far away from be semiclassical.

\begin{figure}[t]
\centering
\includegraphics[width=0.45\textwidth]{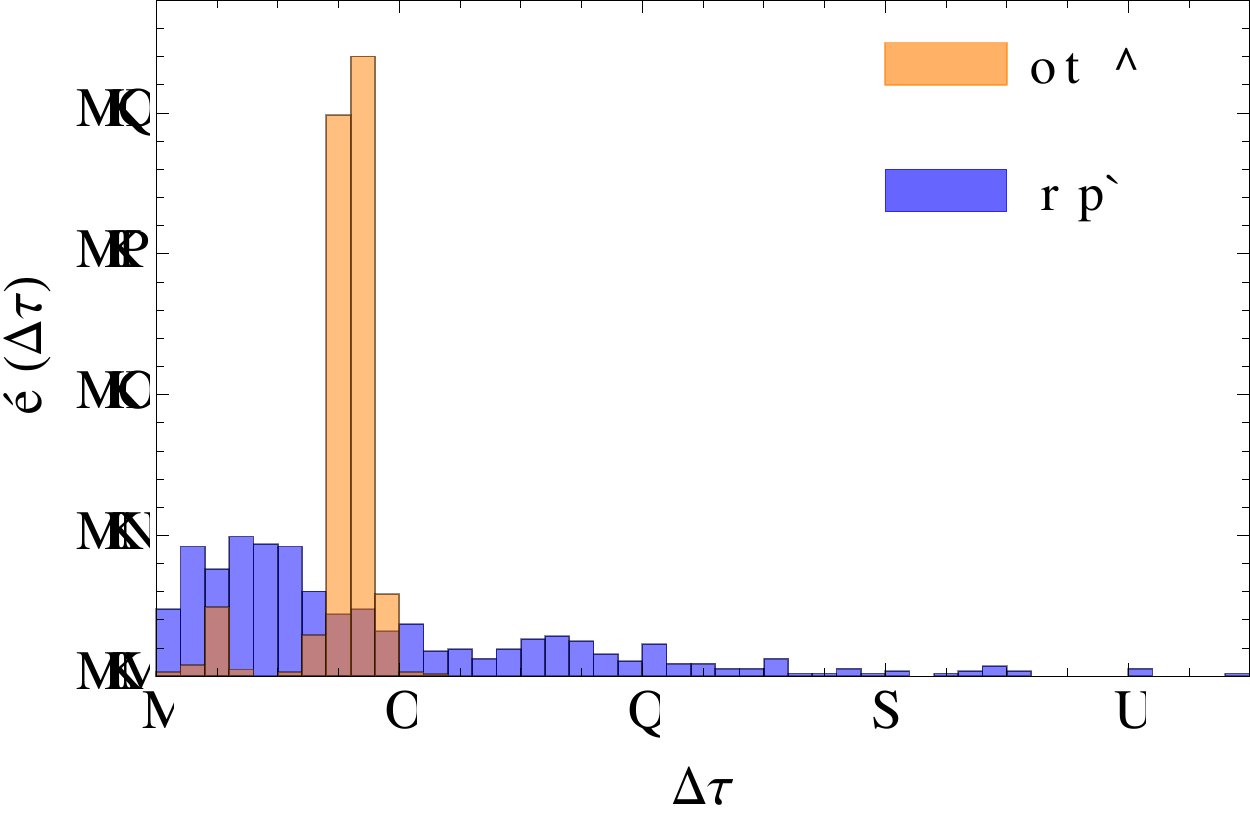}
\caption{(Color online) Probability $p(\Delta \tau)$ of the tunneling times $\Delta \tau$ for $J/\omega=0.5$, $N_0=17$, $\gamma=-1$. The case of $\theta=0$ is shown in orange, $\theta=1$  in blue.}
\label{histo}
\end{figure}

To obtain better insight in the tunneling dynamics, we integrated the numerical diagonalization of \eqref{H} for  $t\in[0,1000]$ and $\Delta \tau$ for $J/\omega=0.5$, $n_0(0)=N_0=17$, $\gamma=-1$. The tunneling times $\Delta \tau_i=\tau_{i+1}-\tau_i$ were defined as the time differences between to consecutive roots of the population imbalance $\rho(J\tau_i)=0$. The probability distribution $p (\Delta \tau)$ is shown in Fig. \ref{histo}. For the case of RWA, two main peaks can be observed, which are caused by  Rabi-oscillations  as well as their collapse and revival. The tunneling time distribution for USC is much broader,  showing that the tunneling times become irregular, although there seam to be reminiscences of the two former peaks with a main peak for lower tunneling times. This behavior is replicated in the structure of the Fourier transform (cf. Fig (3)), where the ultrastrong coupling exhibits, that much more frequencies are involved in the dynamics. 

\paragraph{Discussion.---}

The experimental achievement of the ultra-strong coupling regime in light-matter interactions is a technological revolution in the field solid state quantum optics \cite{us}, quantum information \cite{Peropadre2010, Romero2012} and quantum simulation \cite{tureci, Ballester2012}. However, these experiments have also fundamental theoretical impact, resuscitating old discussions\ \cite{usbef}, about the integrability of the Rabi model or the emergence of chaos, which see the light with new methods to address them\ \cite{Braak2011}. 

The present work studies the presence of chaos in the dynamics of the classical and quantum dimer model. Our approach, which is based on the Heisenberg picture of observables or expectation values, is complementary to alternative studies of quantum chaos, such as the energy level spacing statistics, and provides a very natural connection between the quantum and classical worlds. These techniques can be exported to other problems, such as the Rabi model  \cite{usbef, grigolini}, which may be regarded as a dimer model in which the nonlinearity of one of the wells drops to zero and the other one is raised to infinity, creating a qubit. This particular model has been shown to be integrable in a novel sense \ \cite{Braak2011}, admitting the full classification of energy levels and eigenstates. Compared to the dimer model, it seems that this integrability is enough to support more stable and regular dynamics, and indeed preliminary studies show that the features of quantum chaos in our model are absent in the full Rabi dynamics.

Our works studies the necessary extension of the DNLS beyond the RWA regime.  We have reported on a minimalistic system supporting chaos both in the semiclassical and quantum domains.These versions of the ultra-strong bosonic junction map the semiclassical and quantum limits of two ultra-strongly coupled nonlinear resonators. An implementation of this model consists of two superconducting coplanar waveguide resonators with an embedded Josephson junction or a qubit providing nonlinearity\ \cite{blais1}. Resonator couplings beyond RWA have been experimentally reported\ \cite{Haeberlein2013} and strongly imbalanced states of the microwave resonators can be engineered at will, using, for example, ancilla qubits\ \cite{hofheinz}.

For these models and possible experimental setups our work predicts two main features. The first one is a modification of the transition to self-trapping due to the counterrotating terms. The second feature is the emergence of chaos in the photon number dynamics for attractive bosonic interactions. The physical signature of this novel ultrastrong regime is the  unpredictability of the  tunneling time. We want to emphasize, that in our quantum calculations we did not have to resort to semiclassical or mixed classical-quantum approximation \cite{grigolini}, but have been able to observe chaotic dynamics using a moderate number of excitations, while smooth and purely {\it quantum} tunneling prevails in the deep quantum regime.   Both results can be tested in a coupled-resonator setup, monitoring the field that leaks from the cavity or installing additional qubits that dispersively probe the electromagnetic field. 



 Finally, a natural continuation of this work is the search of similiar features in extended coupled cavity arrays \cite{nlres, Houck2012}, studying the propagation of excitations\ \cite{Peropadre2013}, or the self-trapping dynamics in a many-body setup with dissipation. We believe  that our results are just a few first examples of the rich non-perturbative theoretical landscape opened by the new generation of solid-state experiments.%

\ifcheckpagelimits
\end{document}
\fi

\begin{acknowledgements}
 The authors acknowledge support from the Euro- pean project PROMISCE, Spanish MINECO projects FIS2011-25167 and FIS2012-33022,  Gobierno de Arag\'on (FENOL group) and CAM research consortium QUITEMAD (S2009-ESP-1594).
\end{acknowledgements}

\onecolumngrid
\newpage
\section{Supplement - Physical limits of the parameters}\label{jom}
Starting from the Hamiltonian of two coupled oscillators of masses $m_1$ and $m_2$ with frequencies $\omega_{0,1},\omega_{0,2}$ and coupling strength $C$, 

\bea
H_0&=&\frac{\hat{p}_1^2}{2m_1}+\frac{\hat{p}_2^2}{2m_2}+\frac{m_1\omega_{0,1}^2}{2}\hat{q}_1^2+\frac{m_2\omega_{0,2}^2}{2}\hat{q}_2^2+C(\hat{q}_1-\hat{q}_2)^2\nonumber\\
&=&\frac{\hat{p}_1^2}{2m_1}+\frac{\hat{p}_2^2}{2m_2}-2C\hat{q}_1\hat{q}_2+\Big(\underbrace{\frac{m_1\omega_{0,1}^2}{2}+C}_{\equiv \frac{1}{2}\omega_1^2m_1}\Big)\hat{q}_1^2+\Big(\underbrace{\frac{m_2\omega_{0,2}^2}{2}+C}_{\equiv \frac{1}{2}\omega_2^2m_2}\Big)\hat{q}_2^2\nonumber
\eea

we make the second quantization \mbox{$\hat{q}_k=\sqrt{\frac{\hbar}{2m_k\omega_k}}(\hat{a}_k+\hat{a}_k^+)$} and \mbox{$\hat{p}_k=-i\sqrt{\frac{\hbar m_k\omega_k}{2}}(\hat{a}_k-\hat{a}_k^+)$} for $k={1,2}$. The resulting

\be
H\equiv H_0-\frac{\hbar}{2}(\omega_1+\omega_2)=\hbar \omega_1\hat{a}_1^+\hat{a}_1+\hbar \omega_2\hat{a}_2^+\hat{a}_2-\overbrace{\frac{\hbar C}{\sqrt{m_1m_2\omega_1\omega_2}}}^{\equiv \hbar J}(\hat{a}_1^++\hat{a}_1)(\hat{a}_2^++\hat{a}_2)\nonumber.
\ee

Since at least the initial frequencies should be real and masses positiv, physics requires $m_k,\omega_{0,k}^2\geq 0$. We conclude that 

\be
2 C=2J\sqrt{m_1m_2\omega_1\omega_2}\leq \omega_k^2m_k \quad
\implies J\leq\frac{1}{2}  \textit{Min}\left\lbrace\omega_1\sqrt{\frac{\omega_1 m_1}{\omega_2 m_2}},\omega_2\sqrt{\frac{\omega_2 m_2}{\omega_1 m_1}} \right\rbrace.
\ee

For identical oscillators (symmetric dimer) with $\omega_1=\omega_2=\omega$ and $m_1=m_2=m$, we obtain the limit of "physicality" at $J/\omega\leq 1/2$.

\section{Supplement - Continuation of linear modes}

Starting from eq.$(1)$, we look for the eigenvalues $\dot \psi_k=\lambda \psi_k$. We split $\psi_k=a_k+ib_k$  and obtain the set 

\be
\lambda(a_k+i b_k)= -i \omega (a_k+i b_k) - i \tilde{\gamma} (a_k^2 +b^2_k)(a_k+ib_b)+ i J\left[ (1+\theta)a_{1-k}+i(1-\theta)b_{1-k} \right]\nonumber
\ee

or, ordering into real and imaginary part, 

\be
\lambda\begin{pmatrix}a_1\\a_2\\b_1\\b_2\end{pmatrix}= \mathbf{M} \begin{pmatrix}a_1\\a_2\\b_1\\b_2\end{pmatrix}
\ee

with

\be
\mathbf{M}=\begin{pmatrix}0&0&\omega+\tilde{\gamma}|\psi_1|^2&-J(1-\theta)\\0&0&-J(1-\theta)&\omega+\tilde{\gamma}|\psi_2|^2\\ -\omega-\tilde{\gamma}|\psi_1|^2&J(1+\theta)&0&0\\J(1+\theta)&-\omega-\tilde{\gamma}|\psi_2|^2&0&0\end{pmatrix}.\nonumber
\ee

First, for the linear case $\tilde{\gamma}=0$, we find the four eigenvalues 
\be
\nu=i\lambda=\pm\sqrt{[\omega\pm J(1-\theta)][\omega\pm J(1+\theta)]},
\ee
belonging to a symmetric mode with $\psi_1=\psi_2$ and an antisymmetric mode with $\psi_1=-\psi_2$. We now search for their nonlinear continuations.  Making the Ansatz $|\psi_1|^2=|\psi_2|^2=N/2$, we obtain eqs. $(5)$. At this point we should mention, that the solution of ODE $(1)$ separates no longer into independent stationary mode of the type $\psi_k(t)=\exp(i\nu t)\psi_k$ with time-independent amplitudes $\psi_k$, but has to be constructed like any solution of an homogeneous ODE.

\section{Different parameter regimes}

\begin{figure}[t]
\centering
\includegraphics[width=0.5\textwidth]{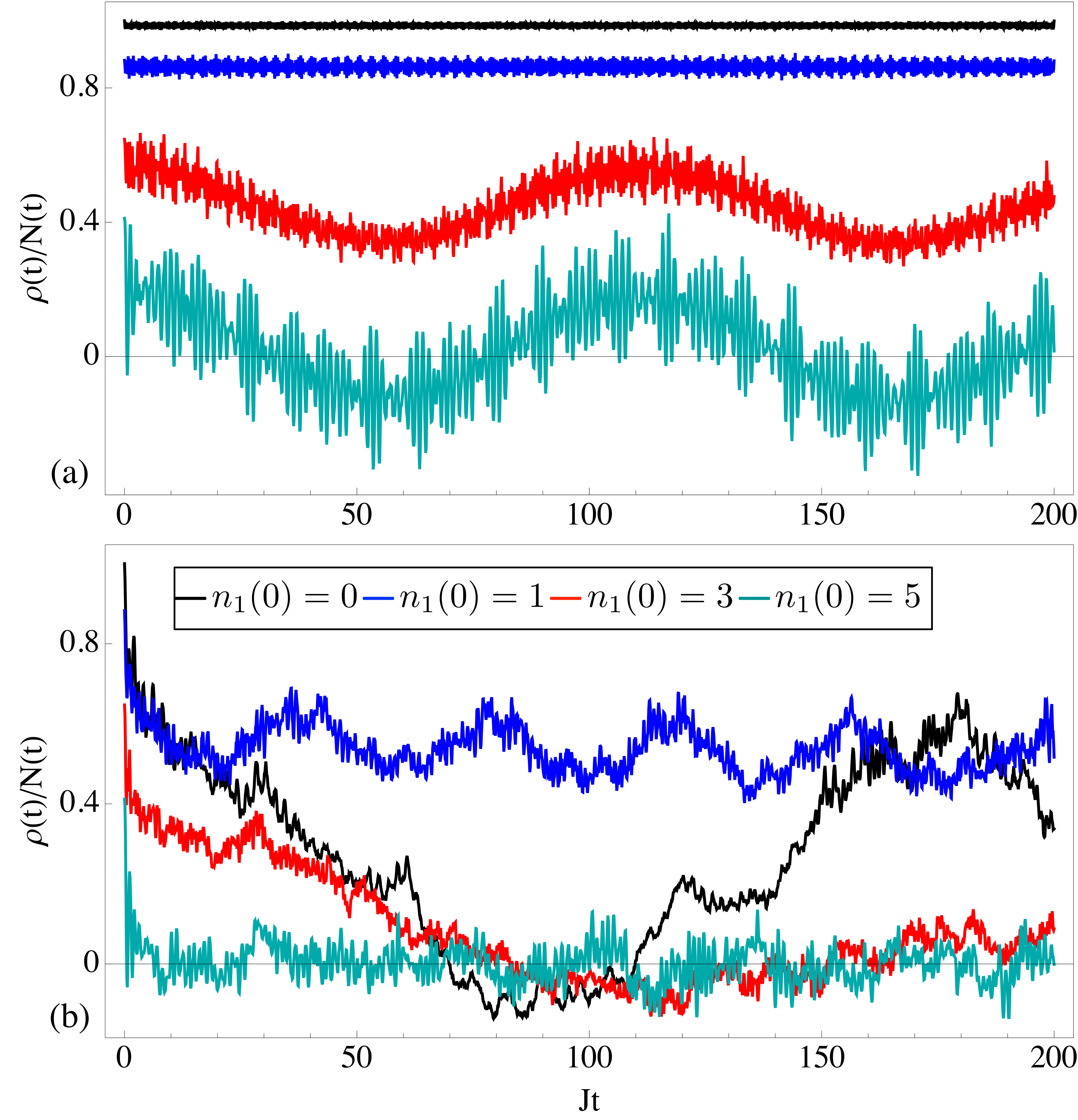}
\caption{(Color online) Quantum dynamics for $J/\omega=0.1$, $N_0=17$, $\rho(t)/N(t)$ vs. $Jt$ is plotted for the USC model at $\gamma=20$ (top figure) and $\gamma =-20$ (bottom). $n_1(0)=0,1,3,5$ (black, blue, red and green, respectively).}
\label{fig4}
\end{figure}

In Fig.\ \ref{fig4} we show that quantum chaos is not only restricted to the ultra-strong coupling value of $J/\omega=0.5$, but  also can  be found  at  $J/\omega=0.1$. In such a case, the absolute value of the nonlinearity  has to be bigger in order to fulfill classical Eq. (6). Therefore, we use $\gamma=20$ for the simulations shown in Fig.~\ref{fig4}. We plot full trajectories of $\rho(t)$ with fixed $N_0=17$ and varying initial imbalance. This is done for positive and negative $\gamma$ in Fig.~\ref{fig4}(a) and (b), respectively. For positive $\gamma$, the most localized initial conditions, $n_1(0)=0,1$ and $3$, preserve localization, but become more and more affected by quasi-periodicity. When the initial state is less localized, $n_1(0)=5$, localization breaks down and we observe incoherent oscillations with $\langle\rho(t)\rangle=0$. For negative $\gamma$, Fig.~\ref{fig4}(b), the picture is different. The most localized initial condition $n_1(0)=0$ has a tunneling probability higher than the less localized initial state $n_1(0)=1$, which decays much slower to $\rho=0$. For further increasing $n_1(0)$, the tunneling time decreases, recovering for $n_1(0)=3$ a similar value like for $n_1(0)=0$ and than dropping further.  

%
%
%
%
%
%


\begin{thebibliography}{43}
\bibitem{footnote} Since $n_{1,2}\in \mathbb{R}$, the negative part of the spectrum is symmetric, $g(-\nu)=g(\nu)$.
\end{thebibliography}
\end{document}